\documentclass[pra,showpacs,floatfix]{revtex4}
\usepackage{amsmath}
\usepackage{amssymb}
\usepackage{graphicx}
\usepackage{color}
\usepackage{subfigure}

\begin{document}

\title{A nonpolynomial Schr\"{o}dinger equation for resonantly absorbing
gratings}
\author{Lior Shabtay and Boris A. Malomed}
\affiliation{Department of Physical Electronics, School of Electrical Engineering,
Faculty of Engineering, Tel Aviv University, Tel Aviv 69978, Israel}
\date{\today }

\begin{abstract}
We derive a nonlinear Schrodinger equation with a radical term, $\pm \sqrt{%
1-|V|^{2}}$, as an asymptotic model of the resonantly absorbing
Bragg reflector (RABR), i.e., a periodic set of thin layers of
two-level atoms, resonantly interacting with the electromagnetic
field and inducing the Bragg reflection. A family of bright solitons
is found, which splits into stable and unstable parts, exactly
obeying the Vakhitov-Kolokolov criterion. The soliton with the
largest amplitude, $\left( |V|\right) _{\max }=1$, is a
``quasi-peakon", i.e., a solution with a discontinuity of the third
derivative at the center. Families of exact cnoidal waves, built as
periodic chains of quasi-peakons, are found too. The ultimate
solution belonging to the family of dark solitons, with the
background level $V=1$, is a dark compacton. Those bright solitons
which are unstable destroy themselves (if perturbed) attaining the
critical amplitude, $|V|~=1$. The dynamics of the wave field around
this critical point is studied analytically, revealing a switch of
the system into an unstable phase, in terms of the RABR model.
Collisions between bright solitons are investigated too. The
collisions between fast solitons are quasi-elastic, while slowly
moving ones merge into breathers, which may persist or perish (in
the latter case, also by attaining $|V|~=1$).
\end{abstract}

\maketitle

\section{Introduction and the model}

The interplay between the resonant reflection of light on Bragg gratings
(BGs) and resonant interaction of light with nanolayers of two-level atoms
(of width $\lesssim 100$ nm, which is much smaller than the wavelength of
light), or with similar active elements, deposited at reflecting layers of
which the BG is built, gives rise to artificial optical media in the form of
\textit{resonantly absorbing Bragg reflectors} (RABRs). They are promising
for fundamental studies and applications \cite{Mantz}-\cite{dip-dip},
including the storage of slow-light pulses \cite{slow}, negative reflection
\cite{negrefl}, and periodically amplifying BGs \cite{amplifyingBG}. The
currently available nanofabrication techniques make the creation of RABRs
with required properties quite feasible \cite{nano}. In particular, the
interplay of the resonant nonlinearity (which gives rise to the self-induced
transparency in uniform media \cite{NM}) with the bandgap spectrum induced
by the BG may give rise to peculiar species of temporal solitons in RABRs
(see Ref. \cite{Progress} for a review, and more recent works \cite%
{recent,Andrei,India}). These solitons belong to the class of \textit{gap
solitons}, whose propagation constant falls into the underlying bandgap. Gap
solitons in fiber Bragg gratings with the uniform Kerr nonlinearity have
been a subject of intensive theoretical studies \cite{Sterke}, and have been
created experimentally, in the form of moving \textit{Bragg solitons} \cite%
{exper}.

Using the rotating-wave approximation, the Maxwell-Bloch equations governing
the transmission of light in the RABR can be reduced to the system of
equations for the scaled slowly varying variables, \textit{viz}., the
amplitude of the electromagnetic field, $\Sigma _{+}$, polarization of the
medium, $P$, and population inversion $w$ of two-level atoms \cite%
{Tomas,Progress}:
\begin{gather}
\left( \Sigma _{+}\right) _{\tau \tau }-\left( \Sigma _{+}\right) _{\zeta
\zeta }=-\eta ^{2}\Sigma _{+}+2i\eta P+2P_{\tau },  \label{Sigma} \\
P_{\tau }=-i\delta P+\Sigma _{+}w,~w_{\tau }=-\mathrm{Re}\ \left( \Sigma
_{+}P^{\ast }\right) ,  \label{w}
\end{gather}%
where $\tau $ and $\zeta $ are the scaled time and coordinate (another
amplitude, $\Sigma _{-}$, is governed by a detached equation). Coefficient $%
\delta $, which may be positive or negative, measures the detuning of the
transition frequency of the two-level atoms from the carrier frequency of
the electromagnetic field, and $\eta $, which is defined to be positive, is
the scaled BG reflectivity.

A straightforward corollary of Eqs. (\ref{w}) is $\partial \left( \left\vert
P\right\vert ^{2}+w^{2}\right) /\partial \tau =0$,~i.e.,$~\left\vert
P\right\vert ^{2}+w^{2}=\mathrm{const}$. The normalization may be set by
fixing $\mathrm{const}=1$, hence $w$ can be eliminated in favor of $P$ \cite%
{Tomas}:
\begin{equation}
w=\pm \sqrt{1-|P|^{2}}.  \label{+-}
\end{equation}%
The stable situation is determined by the condition that, in the absence of
the polarization ($P=0$), the atomic population must be uninverted ($w=-1$),
hence the lower sign must be chosen in Eq. (\ref{+-}). The opposite
situation is possible too, but it is unstable, corresponding to an inverted
population in the absence of the field.

Thus, assuming that the stable branch of square root (\ref{+-}), is used, $%
w=-\sqrt{1-|P|^{2}}$, one arrives at the system of two equations \cite{Tomas}%
,%
\begin{equation}
~\left( \Sigma _{+}\right) _{\tau \tau }-\left( \Sigma _{+}\right) _{\zeta
\zeta }=-\eta ^{2}\Sigma _{+}+2i(\eta -\delta )P-2\sqrt{1-|P|^{2}}\ \Sigma
_{+},  \label{bef1}
\end{equation}%
\begin{equation}
P_{\tau }=-i\delta P-\sqrt{1-|P|^{2}}\ \Sigma _{+}.  \label{bef2}
\end{equation}%
Looking for bright-soliton solutions to Eqs. (\ref{bef1}) and (\ref{bef2})
as $\Sigma _{+}=\exp \left( -i\omega \tau \right) \mathcal{S}(\zeta )$,$%
~P=i\exp \left( -i\omega \tau \right) \mathcal{P}(\zeta )$, solutions for
real functions $\mathcal{S}(\zeta )$ and $\mathcal{P}(\zeta )$ were found in
an implicit analytical form in Ref. \cite{Tomas}. In the general case, these
solitons fall into two distinct bandgaps produced by the linearized version
of the system. Dark-soliton solutions were also studied in Ref. \cite{Tomas}.

The present paper is focused on subfamilies of solitons residing near the
edge of one of the gaps, $\omega =\delta $. First, we aim to shown that, in
an asymptotic approximation valid in this case, Eqs. (\ref{bef1}) and (\ref%
{bef2}) reduce to a single nonpolynomial Schr\"{o}dinger equation (NPSE),
with the nonlinear term in the form of a radical. To this end, solutions are
looked for as
\begin{equation}
\Sigma _{+}=e^{-i\delta \tau }Q^{-1}V(\zeta ,\tau ),~P=-ie^{-i\delta \tau }
\left[ V(\zeta ,t)+R(\zeta ,\tau )\right] ,  \label{V}
\end{equation}%
where $Q\equiv \left( \eta +\delta \right) /2$, $V\left( \zeta ,\tau \right)
$ is assumed to be a slowly varying function of time, in comparison with $%
\exp \left( -i\delta \tau \right) $, and $R$ is a small correction to $V$
required by the self-consistent derivation. Actually, the slow time
dependence in $V$ accounts for a small deviation of the full frequency from $%
\omega =\delta $; in other words, for the purpose of the asymptotic analysis
$\delta $ and $\eta $ may be considered as large parameters, which
corresponds to a far-detuned strongly reflecting RABR. Then, it is a simple
exercise to demonstrate that the self-consistent asymptotic approximation
corresponds to
\begin{equation}
R=\delta ^{-1}\left( iV_{\tau }-Q^{-1}\sqrt{1-|V|^{2}}V\right)  \label{R}
\end{equation}
in Eq. (\ref{V}), and the asymptotic NPSE takes the following form:%
\begin{equation}
iV_{t}+V_{\zeta \zeta }-\epsilon \sqrt{1-|V|^{2}}V=0,  \label{NPSE}
\end{equation}%
where the rescaled time is $t\equiv \delta \left( \eta ^{2}+\delta
^{2}\right) ^{-1}\tau $, and $\epsilon \equiv 2\eta /\delta $. An additional
obvious rescaling of $t$ and $\zeta $ allows one to fix $\epsilon \equiv \pm
1$ for $\delta \gtrless 0$, which is adopted below. It is easy to see that
Eq. (\ref{NPSE}) gives rise to bright- and dark-soliton solutions for $%
\epsilon =+1$ and $-1$, respectively, i.e., for positive and negative values
of the mismatch, $\delta >0$ and $\delta <0$. Equation (\ref{NPSE})
conserves three dynamical invariants, \textit{viz}., the momentum,
Hamiltonian, and norm,
\begin{equation}
N=\int_{-\infty }^{+\infty }\left\vert V(\zeta )\right\vert ^{2}d\zeta .
\label{N}
\end{equation}

Stationary solutions to Eq. (\ref{NPSE}) (in particular, solitons) are
looked for in the usual form,%
\begin{equation}
V\left( \zeta ,t\right) =e^{-i\chi t}W(\zeta ),
\end{equation}%
where real function $W(\zeta )$ satisfies equation
\begin{equation}
\frac{d^{2}W}{d\zeta ^{2}}=-\chi W+\epsilon \sqrt{1-W^{2}}W\equiv -\frac{dU_{%
\mathrm{eff}}}{dW},  \label{W}
\end{equation}%
with effective potential $U_{\mathrm{eff}}=(1/2)\chi W^{2}+(\epsilon
/3)\left( 1-W^{2}\right) ^{3/2}$. Substituting such stationary solutions
back into Eqs. (\ref{V}) and (\ref{R}), one can reproduce the respective
solutions in the framework of the underlying RABR model. In particular, it
is worthy to note that real stationary solutions $W(\zeta )$ correspond,
according to Eqs. (\ref{V}) and (\ref{R}), to \emph{complex} stationary
states of the material polarization, i.e., those with an intrinsic \textit{%
chirp}.

Another noteworthy finding reported below, which directly pertains
to the relation between solutions to the NPSE (\ref{NPSE}) and
solutions of the underlying system of Eqs. (\ref{Sigma}) and
(\ref{w}), is that when, in the course of the dynamical evolution
governed by Eq. (\ref{NPSE}), $\left\vert V\left( \zeta ,t\right)
\right\vert $ attains the critical value, $|V|=1$, i.e., according
to Eq. (\ref{V}), the polarization attains its critical value,
$|P|=1$, the further evolution of the system leads to switching from
the stable branch of relation (\ref{+-}) to the unstable one, i.e., from $w=-%
\sqrt{1-|P|^{2}}$ to $w=+\sqrt{1-|P|^{2}}$. If this happens, the subsequent
use of Eq. (\ref{NPSE}) for the slowly varying field becomes irrelevant,
because the decay of the unstable state may be fast, making it necessary to
get back to the use of the full system (\ref{Sigma}), (\ref{w}) (which is
beyond the scope of this work).

Asymptotic equations for a single slowly varying amplitude can also be
derived near other edges of the two gaps that may be occupied by solitons in
the full system of Eqs. (\ref{Sigma}) and (\ref{w}). However, in other cases
the eventual equation reduces to the usual cubic nonlinear Schr\"{o}dinger
equation (CNLSE). On the other hand, it is relevant to compare the NPSE in
the form of Eq. (\ref{NPSE})\ with the equation which was first derived,
under the same name (NPSE), as the 1D asymptotic form of the
Gross-Pitaevskii equation (GPE)\ for the wave function of a self-attractive
Bose-Einstein condensate (BEC) trapped in a cigar-shaped potential \cite%
{Luca}. In the notation similar to that adopted here, the BEC\ equation
(alias the \textit{Salasnich equation}) is%
\begin{equation}
iV_{t}+V_{\zeta \zeta }+\frac{1-\left( 3/2\right) |V|^{2}}{\sqrt{1-|V|^{2}}}%
V=0.  \label{BEC}
\end{equation}%
Both equations, (\ref{NPSE}) and (\ref{BEC}), give rise to a singularity
when the local amplitude attains the critical value, $|V|=1$. In the case of
Eq. (\ref{BEC}), this singularity leads to the \textit{collapse} of the wave
function, which is a property inherited from the full GPE for the
self-attractive BEC in the three-dimensional space \cite{Luca}. The purport
of the singularity in Eq. (\ref{NPSE}) is demonstrated below: hitting the
critical amplitude, the system switches into the unstable phase, which is
represented, in terms of the underlying RABR model, by square root (\ref{+-}%
) with the upper sign.

Another type of the NPSE was derived in Ref. \cite{Vicente} as the effective
1D reduction of the GPE for the \textit{self-repulsive} BEC. It seems as Eq.
(\ref{NPSE}) with $\epsilon =+1$ and $\sqrt{1-|V|^{2}}$ replaced by $\sqrt{%
1+|V|^{2}}$. Of course, such an equation does not give rise to bright
solitons. Nevertheless, it can generate bright \textit{gap solitons}, if
this nonlinearity is combined with a periodic linear potential (the optical
lattice) \cite{Vicente2}.

The rest of the paper is organized as follows. In Section II, we report
analytical results for bright and dark solitons in Eq. (\ref{NPSE}) with $%
\epsilon =+1$ and $-1$, respectively, as well as for cnoidal waves in the
former case, and for the switch of the system into the unstable phase, in
both cases. In the same section we also report results of simulations
confirming the stability and instability of the bright solitons, as
predicted in the analytical form by means of the Vakhitov-Kolokolov (VK)
criterion. Simulations of 2-soliton states and collisions between moving
bright solitons are reported in Section III. The paper is concluded by
Section IV.

\section{Analytical results}

\subsection{Bright solitons}

Solutions to stationary equation (\ref{W}) can be represented by means of
the formal energy integral,
\begin{equation}
\left( \frac{dW}{d\zeta }\right) ^{2}+\chi W^{2}+\frac{2\epsilon }{3}\left(
1-W^{2}\right) ^{3/2}=\mathrm{const}.  \label{soliton}
\end{equation}

For bright solitons, with $W\left( |\zeta |~=\infty \right) =0$, which
correspond to $\epsilon =+1$, as said above, and, accordingly, $\mathrm{const%
}=2/3$ in Eq. (\ref{soliton}), $W(\zeta )$ attains its maximum value ($A$)
at the center of the soliton, where $dW/d\zeta $ vanishes. Therefore,
amplitude $A$ can be found by setting $dW/d\zeta =0$ in Eq. (\ref{soliton})
with $\epsilon =+1$:%
\begin{equation}
A^{2}=\frac{3}{2}\left[ 1-\frac{3}{4}\chi ^{2}-\sqrt{\left( 1-\frac{3}{4}%
\chi ^{2}\right) ^{2}-\frac{4}{3}\left( 1-\chi \right) }\right] .
\label{A^2}
\end{equation}%
While the bandgap where bright solitons may reside is, formally,
semi-infinite in the framework of Eq. (\ref{NPSE}): $\chi <1$, the solitons
actually exist in a finite interval, $2/3<\chi <1$, in which squared
amplitude (\ref{A^2}) varies from $1$ to $0$. At the limit point of $\chi
=2/3$, the soliton solution can be found in an explicit form:%
\begin{equation}
W_{\chi =2/3}(\zeta )=2\tanh \left( \frac{|\zeta |}{\sqrt{3}}+\ln \left(
\sqrt{2}+1\right) \right) \mathrm{sech}\left( \frac{|\zeta |}{\sqrt{3}}+\ln
\left( \sqrt{2}+1\right) \right) .  \label{peakon}
\end{equation}%
The expansion of solution (\ref{peakon}) around zero is
\begin{equation}
W_{\chi =2/3}(\zeta )=1-\left( 1/3\right) \zeta ^{2}+\left( 1/6\right) \sqrt{%
2/3}\left\vert \zeta \right\vert ^{3}+\mathcal{O}\left( \zeta ^{4}\right) .
\label{expansion}
\end{equation}%
As seen from here, a peculiarity of this solution is that, while both $%
W(\zeta )$ and its first two derivatives are continuous at $\zeta =0$, the
third derivative, $d^{3}W/d\zeta ^{3}$, suffers a discontinuity, jumping
from $-\sqrt{2/3}$ to $+\sqrt{2/3}$ as $\zeta $ crosses zero. In this sense,
this exact solution may be called a \textit{quasi-peakon}, usual peakons
being solitons with a jump of the first derivative at the center \cite%
{peakon}.

\begin{figure}[b]
\centering
\includegraphics[width=0.60\textwidth]{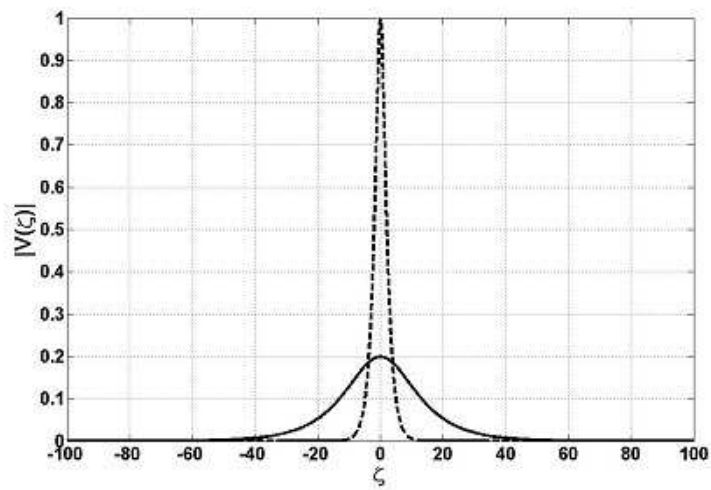}
\caption{The solitonic profiles corresponding to $\protect\chi =0.67$ and $%
\protect\chi =0.99$ (dashed and solid curves, respectively). The numerically
found solutions are indistinguishable from, severally, the exact
``quasi-peakon" solution (\protect\ref{peakon}) found for $%
\protect\chi =2/3$, and the approximate solution (\protect\ref{NLS}) which
is relevant for $1-\protect\chi \ll 1$.}
\label{fig1}
\end{figure}

In the limit of $1-\chi \rightarrow 0$, the smallness of amplitude (\ref{A^2}%
) suggests expanding the radical in Eq. (\ref{NPSE}), which reduces the
equation to the usual CNLSE, and the soliton solutions, accordingly, take
the following form:%
\begin{equation}
V=2\sqrt{1-\chi }e^{-i\chi t}\mathrm{sech}\left( \sqrt{1-\chi }\zeta \right)
.  \label{NLS}
\end{equation}%
Solitons close to those given by Eqs. (\ref{peakon}) and (\ref{NLS}) are
displayed in Fig. \ref{fig1}.

\begin{figure}[b]
\centering$%
\begin{array}{cc}
\includegraphics[width=0.40\textwidth]{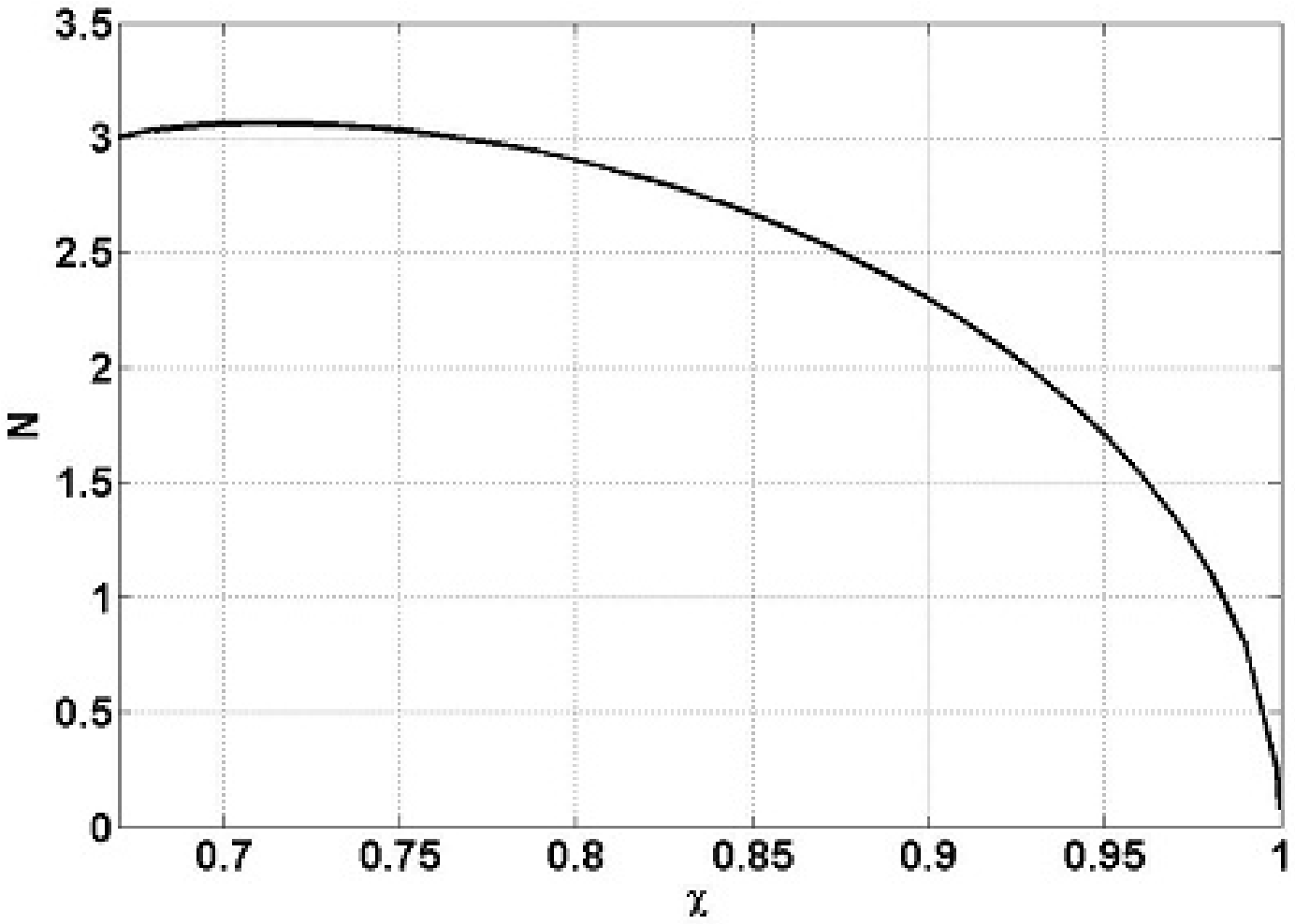} & %
\includegraphics[width=0.40\textwidth]{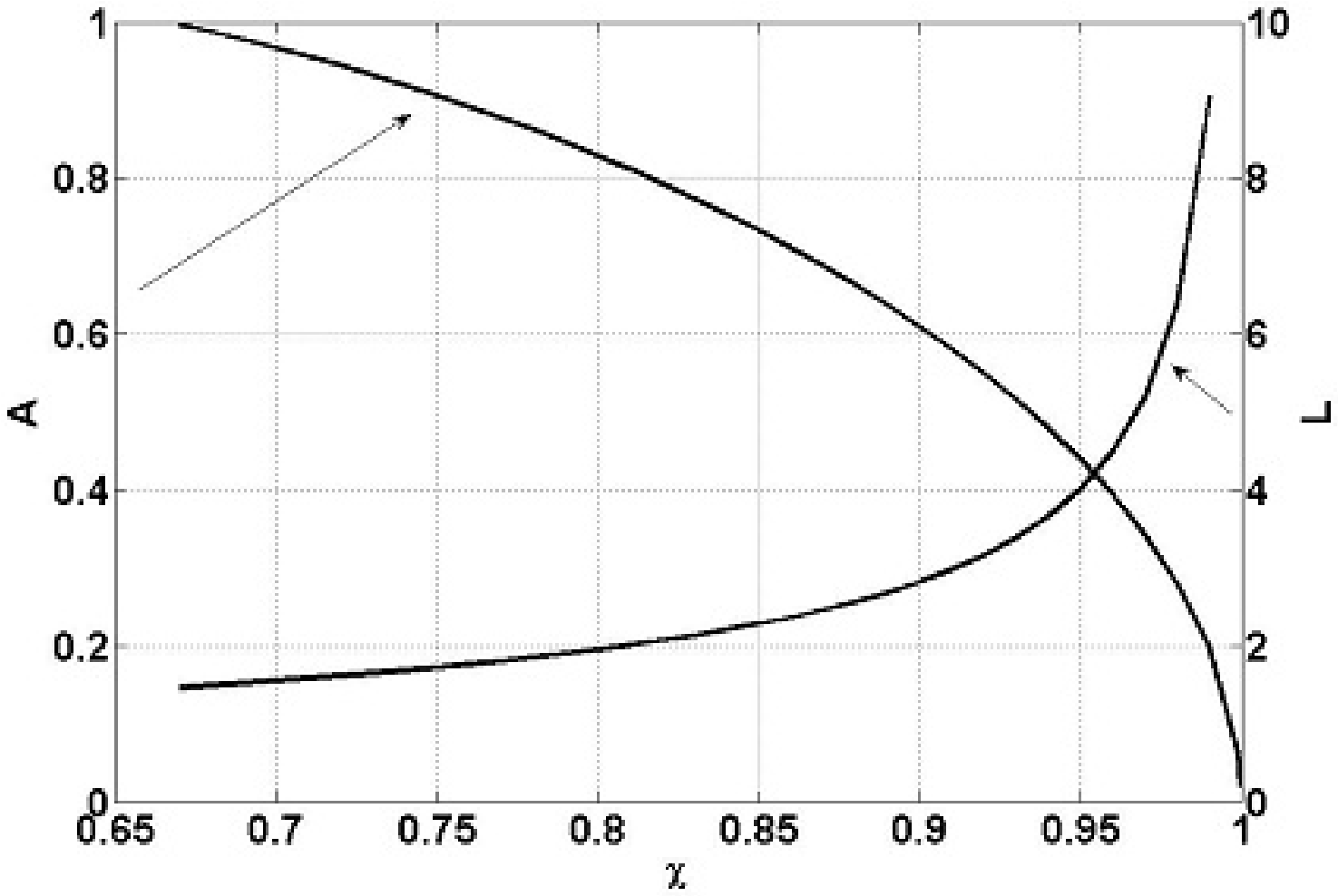} \\
\mathrm{(a)} & \mathrm{(b)}%
\end{array}%
$%
\caption{The norm (a) and amplitude and width (b) of the soliton versus its
intrinsic frequency $\protect\chi $. According to the VK criterion, the
solitons are stable at $dN/d\protect\chi <0$.}
\label{fig2}
\end{figure}

The stability of the solitons can be predicted by means of the
Vakhitov-Kolokolov (VK) criterion, $dN/d\chi <0$, where $N$ is the norm
defined as per Eq. (\ref{N}) \cite{VK}. The numerically calculated curve $%
N(\chi )$ for the entire soliton family is displayed in Fig. \ref{fig2},
along with the amplitude, $A(\chi )$, as given by analytical expression (\ref%
{A^2}) (which completely coincides with its numerical counterpart), and the
soliton's width, $L(\chi )$, defined as $L^{2}\equiv N^{-1}\int_{-\infty
}^{+\infty }\zeta ^{2}W^{2}(\zeta )d\zeta $. It is seen in Fig. \ref{fig2}%
(a) that the VK criterion predicts the stability of the solitons in interval%
\begin{equation}
\chi _{\mathrm{cr}}\approx 0.7120<\chi <1\text{,}  \label{stab}
\end{equation}%
and instability in the remaining part of the existence region, $2/3<\chi
<\chi _{\mathrm{cr}}$.

\begin{figure}[b]
\centering$%
\begin{array}{c}
\includegraphics[width=0.65\textwidth]{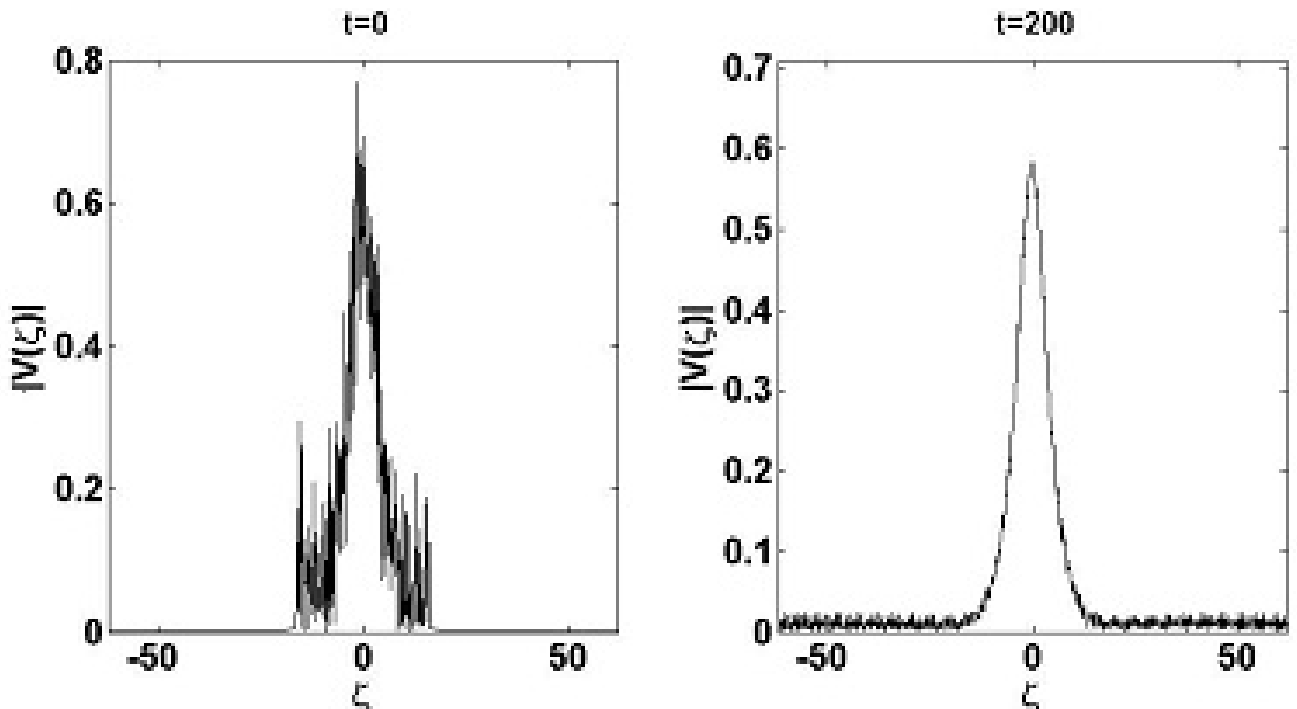} \\
\mathrm{(a)} \\
\includegraphics[width=0.65\textwidth]{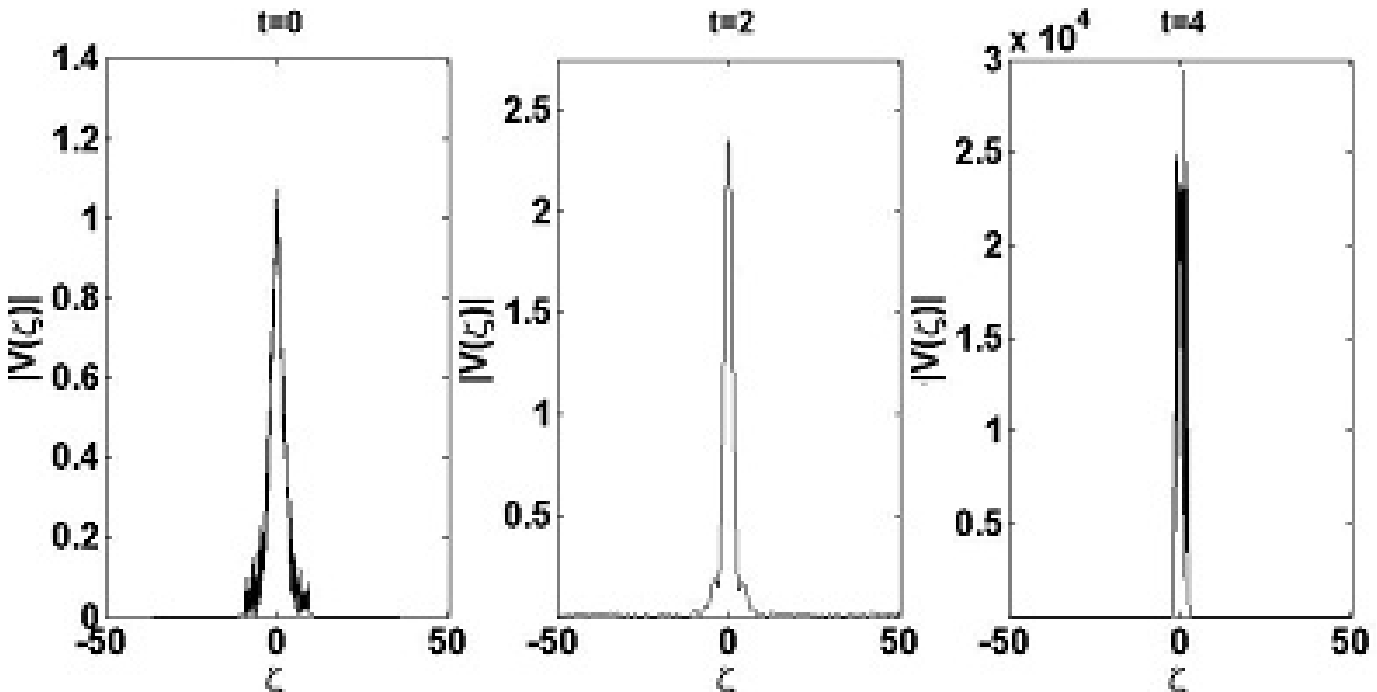} \\
\mathrm{(b)}%
\end{array}%
$%
\caption{(a) Self-cleaning of a stable soliton with $\protect\chi =0.9$, to
which a strong random perturbation was added at $t=0$. (b) The evolution of
a slightly perturbed soliton with $\protect\chi =0.69$, which belongs to the
VK-unstable subfamily. The crossing of the critical level, $|V|=1$, implies
the transition into the unstable phase. The subsequent blowup indicates the
loss of the system's stability.}
\label{fig3}
\end{figure}

Direct simulations of the evolution of perturbed bright solitons, performed
in the framework of Eq. (\ref{NPSE}), corroborate this prediction: strong
perturbations added to VK-stable solitons gradually fade, leaving the
soliton intact, as shown in Fig. \ref{fig3}(a). On the other hand, the
evolution of perturbed VK-unstable solitons does not lead to their immediate
destruction. Rather, as shown in Fig. \ref{fig3}(b), the amplitude of the
soliton grows, crossing the critical level, $|V|~=1.$ Then, formal
continuation of the simulations shows a blowup [see, e.g., the panel
corresponding to $t=4$ in Fig. \ref{fig3}(b)], which is a manifestation of
the fact that the model as a whole becomes unstable after hitting the
critical level, see below.

\subsection{Cnoidal waves}

In addition to the bright solitons, a subfamily of exact periodic solutions
in the form of \textit{cnoidal waves} (expressed in terms of the Jacobi's
elliptic functions) can also be found from energy equation (\ref{soliton})
with $\epsilon =+1$, by setting $\mathrm{const}$ $=\chi $ on its right-hand
side. First, in the case of

\begin{equation}
0<\chi <2/3,  \label{case1}
\end{equation}%
the period of the cnoidal solution is defined as an interval of coordinate $%
\zeta $ in which a continuously varying elliptic function, $\mathrm{sn}%
\left( \zeta /\sqrt{3},k\right) $ with modulus
\begin{equation}
k=\sqrt{\left( 1/2\right) \left[ 1+(3/2)\chi \right] },  \label{k}
\end{equation}%
takes values that are not too small:%
\begin{equation}
\left( \sqrt{2}k\right) ^{-1}\equiv \left( 1+3\chi /2\right) ^{-1/2}\leq
\mathrm{sn}\left( \zeta /\sqrt{3},k\right) \leq 1.  \label{>}
\end{equation}%
In this interval, the solution is
\begin{equation}
W(\zeta )=2k~\mathrm{sn}\left( \zeta /\sqrt{3},k\right) \mathrm{dn}\left(
\zeta /\sqrt{3},k\right) ,  \label{cn1}
\end{equation}%
where \textrm{dn} is the other standard elliptic function, and the entire
solution is built as a chain of the so defined periods. Note that $k$ given
by Eq. (\ref{k}) takes values $1/\sqrt{2}<k<1$ if $\chi $ belongs to region (%
\ref{case1}). As follows from Eq. (\ref{cn1}) and (\ref{>}), at junction
points between adjacent periods, where the left inequality in Eq. (\ref{>})
turns into the equality, the solution attains the critical value, $W=1$, and
its derivative vanishes, $dW/d\zeta =0$, hence the matching at the junctions
is continuous for both $W(\zeta )$ and $dW/d\zeta $, as well as for $%
d^{2}W/d\zeta ^{2}$ [according to Eq. (\ref{W}), $d^{2}W/d\zeta ^{2}=-\chi $
at $W=1$]. Only the third derivative is discontinuous at the junction,
jumping between $-\sqrt{\chi }$ and $+\sqrt{\chi }$. In this sense, these
cnoidal solutions are built as periodic chains of \textit{quasi-peakons},
cf. the exact soliton solution (\cite{peakon}) of the same type. Note also
that the bright solitons do not exist in the entire interval (\ref{case1}),
i.e., the existence regions of this type of the cnoidal waves and solitons
are separated.

Exact solutions for cnoidal waves take a different form at $\chi >2/3$
(recall the solitons exist in the region of $2/3\leq \chi <1$, i.e., the
cnoidal waves may coexist with the solitons in this case, although the
cnoidal solutions exists also at $\chi >1$, where the solitons cannot be
found). In this case, the solution is built of elliptic functions $\mathrm{sn%
}$ and $\mathrm{cn}$ with modulus
\begin{equation}
k=\sqrt{2\left[ 1+(3/2)\chi \right] ^{-1}},  \label{k2}
\end{equation}%
cf. Eq. (\ref{k}). Note that expression (\ref{k2}) is the inverse of (\ref{k}%
), and it takes values $k<1$ for $\chi >2/3$. The period of the cnoidal
solution is now defined by inequalities%
\begin{equation}
1/\sqrt{2}\leq \mathrm{sn}\left( \sqrt{(1/6)\left[ 1+(3/2)\chi \right] }%
\zeta ,k\right) \leq 1,  \label{>2}
\end{equation}%
cf. Eq. (\ref{>}), and the solution is
\begin{equation}
W(\zeta )=2~\mathrm{sn}\left( \sqrt{(1/6)\left[ 1+(3/2)\chi \right] }\zeta
,k\right) \mathrm{cn}\left( \sqrt{(1/6)\left[ 1+(3/2)\chi \right] }\zeta
,k\right) .  \label{cn2}
\end{equation}%
At junction points between adjacent periods, $W$ again attains the
critical value, $W=1$, the first derivative \ vanishes, $dW/d\zeta
=0$, the second derivative is continuous, taking value
$d^{2}W/d\zeta ^{2}=-\chi $, while the third derivative jumps
between values $\pm \sqrt{\chi }$[cf. the jump of the third
derivative in expansion (\ref{expansion}) for the ``quasi-peakon"].

Actually, the quasi-peakon solution (\ref{peakon}) corresponds to the limit
form of both cnoidal families at the border between them, $\chi =2/3$. We
stress that the above exact cnoidal-wave solutions, which depend on the
single parameter, $\chi $, represent only particular cases of a general
family of periodic solutions, which depend on two parameters, $\chi $ and $%
\mathrm{const}$ in Eq. (\ref{soliton}) (and cannot be expressed in terms of
the Jacobi's elliptic functions).

\subsection{Dark solitons}

An obvious condition necessary for the stability of dark solitons is the
modulational stability of the CW (continuous-wave) states,
\begin{equation}
V_{\mathrm{CW}}=e^{-i\chi t}\sqrt{1-\chi ^{2}},  \label{CW}
\end{equation}%
with $\chi $ taking values $-1<\chi <0$. A straightforward analysis
demonstrates that all CWs (\ref{CW}) are indeed stable, if $\epsilon =-1$ in
Eq. (\ref{NPSE}).

Solutions for dark solitons, $V\left( \zeta ,t\right) =e^{-i\chi t}W(\zeta )$%
, which approach CW (\ref{CW}) with frequency $\chi $ at $\zeta \rightarrow
\pm \infty $, can be found as solutions to Eq. (\ref{soliton}) with $%
\epsilon =-1$ and $\mathrm{const}=\chi -(1/3)\chi ^{3}$. A straightforward
analysis demonstrates that, in the limit case of $\chi =0$, the solution
takes a peculiar form of a \textit{compacton-shaped }dark soliton:%
\begin{equation}
W(\zeta )=\left\{
\begin{array}{c}
\mathrm{sn}\left( \zeta /\sqrt{3},1/\sqrt{2}\right) \sqrt{2-\mathrm{sn}%
^{2}\left( \zeta /\sqrt{3},1/\sqrt{2}\right) },~\mathrm{at}~|\zeta |<\sqrt{3}%
K\left( 1/\sqrt{2}\right) , \\
1,~\mathrm{at}~|\zeta |~\geq \sqrt{3}K\left( 1/\sqrt{2}\right) ,%
\end{array}%
\right.   \label{comp}
\end{equation}%
where the modulus of $\mathrm{sn}$ is $1/\sqrt{2}$, and $K\left( 1/\sqrt{2}%
\right) $ is the corresponding value of the complete elliptic integral of
the first kind. Previously, dark compactons were reported in several
discrete models \cite{compacton}, but we are not aware of solutions similar
to that given by Eq. (\ref{comp}) in continual models.

\subsection{The switch into the unstable phase}

The crossing of the critical amplitude level, $|V|~=1$, by an evolving
solution can be studied in an analytical form too. To this end, a
near-critical solution is looked for as
\begin{equation}
V\left( \zeta ,t\right) =1-\left[ v_{1}\left( \zeta ,t\right) +iv_{2}\left(
\zeta ,t\right) \right] ,  \label{v1v2}
\end{equation}%
where functions $v_{1}$ and $v_{2}$ are real and imaginary parts of small
perturbations around $V=1$, i.e., $v_{1,2}^{2}\ll 1.$The substitution of
this expression into Eq. (\ref{NPSE}) and a straightforward asymptotic
expansion yields the following equations:%
\begin{eqnarray}
\left( v_{2}\right) _{t}-\left( v_{1}\right) _{\zeta \zeta }-\epsilon \sqrt{%
2v_{1}} &=&0,  \label{vv1} \\
\left( v_{1}\right) _{t}+\left( v_{2}\right) _{\zeta \zeta } &=&0.
\label{vv2}
\end{eqnarray}%
Solutions to Eq. (\ref{vv1}) make sense if $v_{1}\left( \zeta ,t\right) $
does not become negative.

\subsubsection{The case of $\protect\epsilon =-1$}

First, we consider this issue for Eq. (\ref{NPSE}) with $\epsilon =-1$,
which admits stable CW states. In this case, a family of \emph{exact}
solutions to asymptotic equations (\ref{vv1}) and (\ref{vv2}) can be looked
for in the following form:%
\begin{equation}
v_{1}\left( \zeta ,t\right) =(1/2)\left( \beta \zeta ^{2}-at\right)
^{2},~v_{2}\left( \zeta ,t\right) =(1/2)\left[ bt^{2}+ct\zeta ^{2}+\left(
\gamma /2\right) \zeta ^{4}\right] ,  \label{ansatz}
\end{equation}%
with constants $a,\beta ,b,c,\gamma $. Here $t=0$ is defined as the moment
of time at which $v_{1}$ vanishes for the first time, i.e., the critical
value $|V|~=1$ is attained, in the framework of the asymptotic
approximation. Note that the choice of the ansatz for $v_{1}$ in the form of
the full square in expression (\ref{ansatz}) guarantees that $v_{1}\left(
t,\zeta \right) $ remains positive, as it must be. The substitution of the
ansatz into Eqs. (\ref{vv1}) and (\ref{vv2}) gives rise to the following
relations:
\begin{equation}
b=\left( 1-2\beta \right) a,~c=-a^{2},~\gamma =(1/3)a\beta ,~a=\sqrt{2\beta
\left( 1-6\beta \right) },  \label{a}
\end{equation}%
where $\beta $ remains an arbitrary parameter, taking values $0<\beta <1/6$.

Ansatz (\ref{ansatz}), subject to conditions (\ref{a}), yields an exact
solution to Eqs. (\ref{vv1}) and (\ref{vv2}) with $\epsilon =-1$, provided
that $\sqrt{2v_{1}\left( \zeta ,t\right) }$ is realized, when the ansatz is
substituted into Eq. (\ref{vv1}), as $\beta \zeta ^{2}-at$, but \emph{not}
as $\left\vert at-\beta \zeta ^{2}\right\vert $ (the latter expression
cannot provide for a solution). An explicit form of $\sqrt{1-|V|^{2}}$, as
given by Eqs. (\ref{v1v2}), (\ref{ansatz}), and (\ref{a}), with regard to
the above-mentioned realization of $\sqrt{2v_{1}\left( \zeta ,t\right) }$,
is
\begin{equation}
\sqrt{1-|V|^{2}}\approx \sqrt{2v_{1}\left( \zeta ,t\right) }=\beta \zeta
^{2}-\sqrt{2\beta \left( 1-6\beta \right) }t.  \label{sqrt}
\end{equation}%
As follows from Eq. (\ref{sqrt}), $\sqrt{1-|V|^{2}}$ does not vanish (i.e., $%
|V|~<1$ holds) at $t<0$, which is the \textit{pre-critical} stage of the
evolution. At the \textit{critical moment} of time, $t=0$, $\sqrt{1-|V|^{2}}$
vanishes at point $\zeta =0$. Then, as seen from Eq. (\ref{vv1}), at $t>0$
(at the \textit{post-critical} stage) $\sqrt{1-|V|^{2}}$ vanishes at two
points, $\zeta _{0}(t)=\pm \sqrt{2\left( \beta ^{-1}-6\right) t}$. In the
\textit{instability domain} between these points,
\begin{equation}
-\sqrt{2\left( \beta ^{-1}-6\right) t}<\zeta <+\sqrt{2\left( \beta
^{-1}-6\right) t},  \label{domain}
\end{equation}%
which emerges at $t=0$ and expands with time as $\sqrt{t}$, expression (\ref%
{sqrt}), i.e., eventually, square root (\ref{+-}) [see Eq. (\ref{V})],
switches from the stable (lower) branch to the unstable (upper) one. Further
evolution of the system is expected to be strongly affected by the presence
of the instability domain, and should be studied by means of direct
simulations of the full system of Eqs. (\ref{Sigma}) and (\ref{w}), which is
is beyond the scope of this work.

\subsubsection{The case of $\protect\epsilon =+1$}

In the case when the bright solitons exist, i.e., $\epsilon =+1$, the
critical level, $|V|~=1$, is attained at the center of \textit{quasi-peakon}
(\ref{peakon}). In this case, it is not possible to find an exact solution
to Eqs. (\ref{vv1}) and (\ref{vv2}) describing the crossing of $|V|~=1$ by
a\ perturbed peakon, unlike the above solution given by Eqs. (\ref{ansatz})
and (\ref{a}). However, taking into regard expansion (\ref{expansion}) of
the quasi-peakon around its center, and the fact that its frequency is $\chi
=2/3,$ an approximate nonstationary solution can be sought for, at small $%
|\zeta |$ and $|t|$, as%
\begin{eqnarray}
v_{1} &=&\left[ \left( 1/\sqrt{3}\right) |\zeta |~-at\right] ^{2}+O\left(
\zeta ^{4}\right) ,  \label{cusp} \\
v_{2} &=&(2/3)t+\sqrt{2/3}t|\zeta |~-\left( 1/\sqrt{2}\right) at^{2}+O\left(
|\zeta |^{3},t\zeta ^{2}\right) ,  \notag
\end{eqnarray}%
where $a>0$ is an arbitrary constant, and the square root is realized as
follows:%
\begin{equation}
\sqrt{1-|V|^{2}}\approx \sqrt{2v_{1}\left( \zeta ,t\right) }\approx \left(
\sqrt{2/3}\right) |\zeta |~-\sqrt{2}at,  \label{sqrt2}
\end{equation}%
cf. Eq. (\ref{sqrt}). As seen from Eq. (\ref{cusp}), at $t\rightarrow -0$
this approximate solution describes a small perturbation in the form of a
\textit{cuspon} introduced at the central point, $\zeta =0$. At $t>0$, Eq. (%
\ref{sqrt2}) demonstrates that the square root switches into the \emph{%
unstable branch} in the respective instability domain, $|\zeta |~<\sqrt{3}at$%
, which expands linearly with $t$, cf. Eq. (\ref{domain}). As said above,
Eq. (\ref{NPSE}) cannot be used after the switching into the unstable phase.

\section{Dynamics of bright solitons}

\begin{figure}[b]
\centering$%
\begin{array}{cc}
\includegraphics[width=0.60\textwidth]{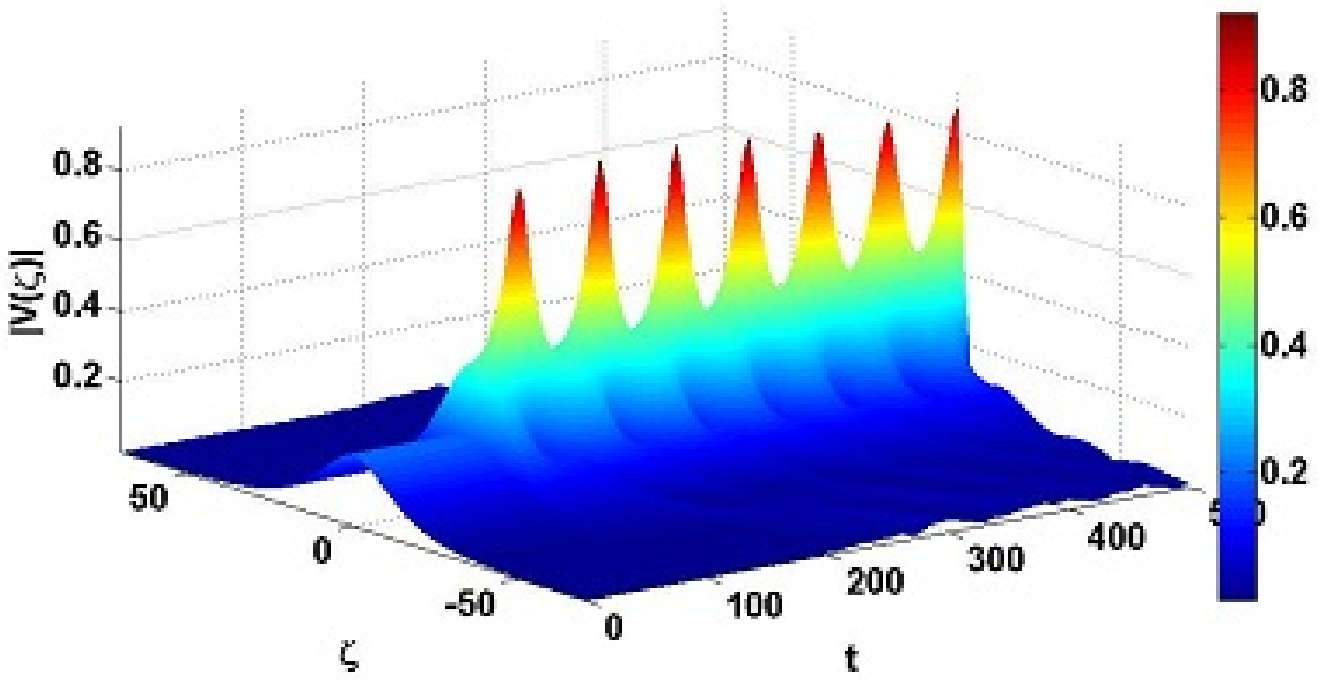} & %
\includegraphics[width=0.45\textwidth]{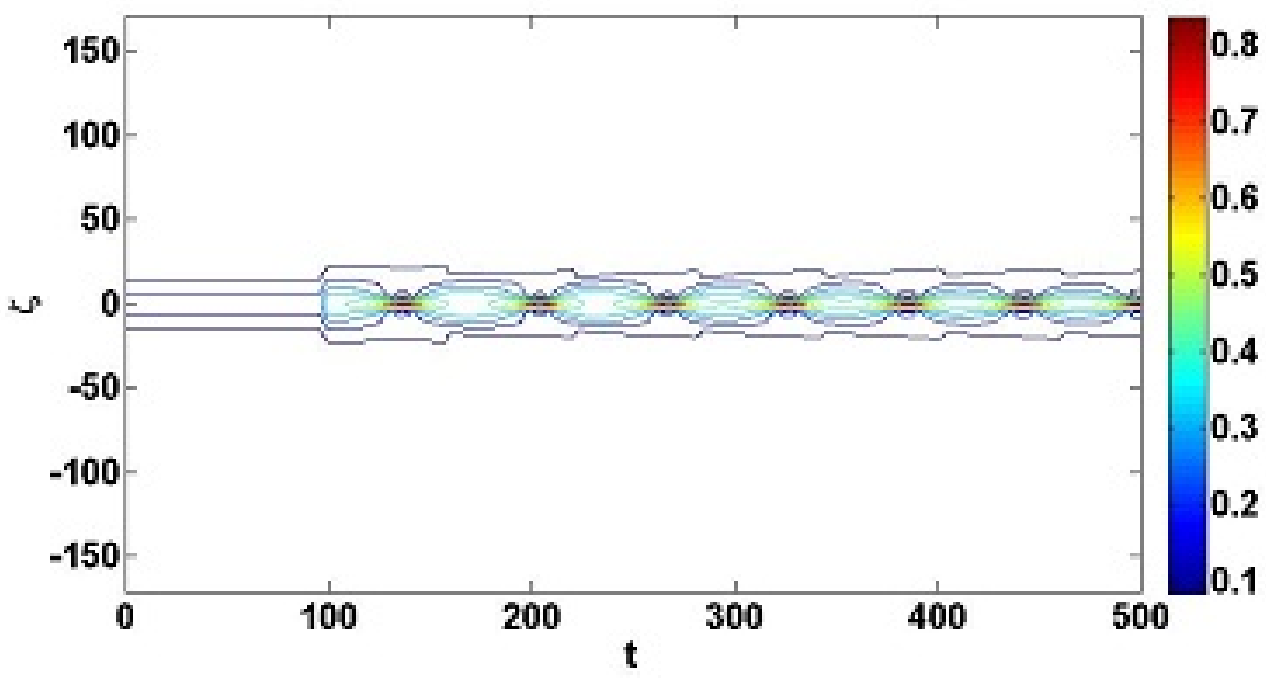} \\
\mathrm{(a)} & \mathrm{(b)}%
\end{array}%
$%
\caption{(Color online) An example of the breather generated by the
2-soliton initial condition in the case of $\protect\chi =0.99$ (the
fundamental-soliton solution is suddenly multiplied by $2$ at $t=100$): (a)
the 3D image; (b) the contour plot in the $\left( t,\protect\zeta \right) $
plane.}
\label{fig4}
\end{figure}

It was mentioned above that, in the limit of $1-\chi \rightarrow 0$, Eq. (%
\ref{NPSE}) goes over into the CNLSE. A well-known peculiarity of the latter
(integrable) equation is the existence of higher-order soliton solutions
(breathers), the simplest one, \textit{2-soliton}, being obtained by the
multiplication of a fundamental soliton by $2$ at the initial moment \cite%
{2soliton}. This circumstance suggests to simulate the evolution of so
produced double pulses in the framework of Eq. (\ref{NPSE}), if the initial
fundamental soliton is taken with $1-\chi $ small enough. A typical example
is displayed in Fig. \ref{fig4}, which demonstrates that, in this case, Eq. (%
\ref{NPSE}) indeed supports robust breathers of the 2-soliton type, although
some emission of radiation is observed too.

\begin{figure}[b]
\centering
\includegraphics[width=1.15\textwidth]{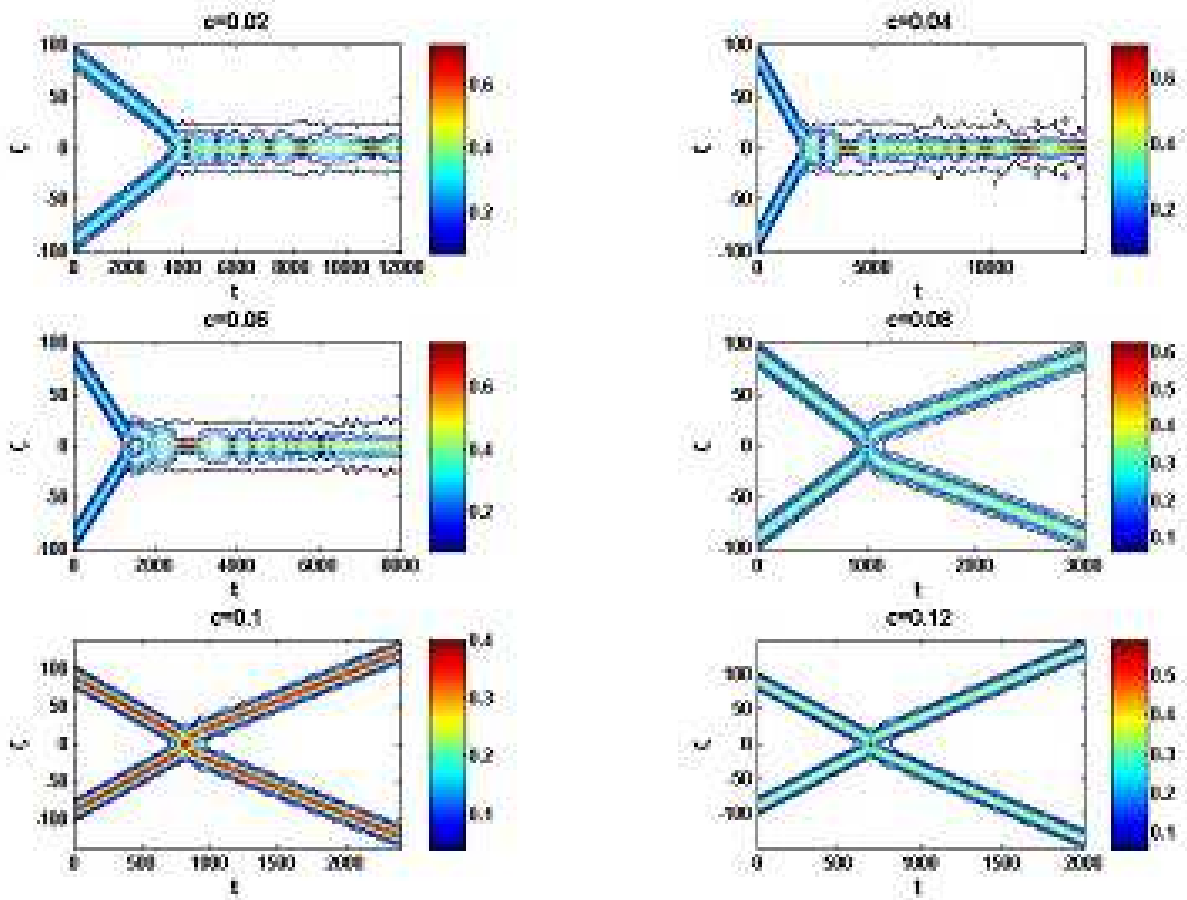}
\caption{(Color online) Outcomes of collisions between identical stable
solitons with $\protect\chi =0.96$, observed with a gradual increase of the
collision velocity, $c$. A transition from the merger to quasi-elastic
collisions is observed at $c=c_{\min }\approx 0.07$.}
\label{fig5}
\end{figure}

Collisions between moving solitons is another natural dynamical problem \cite%
{RMP}. The corresponding simulations were carried out, as usual, by taking a
pair of far separated identical stable solitons [those with $\chi >\chi _{%
\mathrm{cr}}$, see Eq. (\ref{stab})] and setting them in motion with
velocities $\pm c$ by the application of the ``kick", i.e., multiplying each
soliton by $\exp \left( \pm ic\zeta /2\right) $. Figure \ref{fig5}
demonstrates that, as might be expected, the collisions are quasi-elastic if
$c$ exceeds a certain minimum velocity, $c_{\min }$, while, at $c<c_{\min }$%
, the colliding solitons merge into a single pulse which features irregular
intrinsic oscillations (as seen from the figure, the quasi-elastic
collisions may result in a change of the velocities).

In fact, the latter outcome is observed only for $\chi \geq 0.96$, i.e.,
close enough to the CNLSE limit. At $\chi <0.96$, the merger produces pulses
whose amplitude exceeds the critical level, $|V|~=1$ (not shown here in
detail), which is followed by the switch of the pulse into the unstable
phase, as outlined above .

\begin{figure}[b]
\centering
\includegraphics[width=0.60\textwidth]{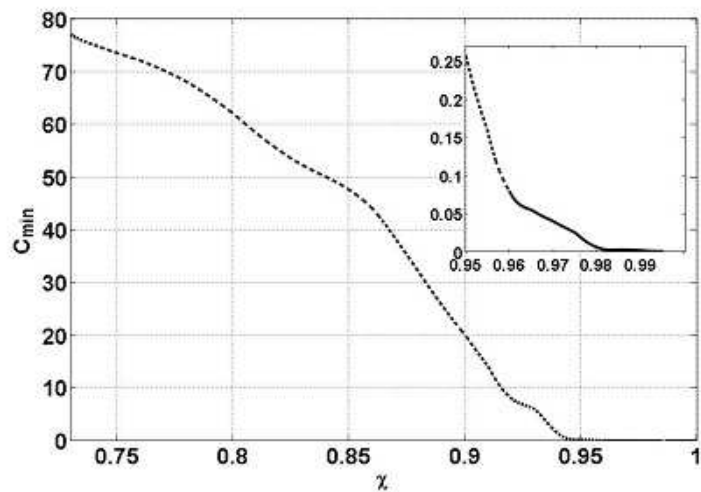}
\caption{The velocity separating quasi-elastic collisions and the merger of
identical in-phase solitons, versus the intrinsic frequency of the soliton, $%
\protect\chi $. The inset zooms the region of small $1-\protect\chi $. The
solid and dashed portions of the curve correspond, respectively, to the
merger of the colliding solitons into breathers remaining in the stability
domain (at $|V|~<1$), and to the case when the pulses produced by the merger
(at $c<c_{\min }$) attain values $|V|~=1$ and thus switch into the unstable
phase.}
\label{fig6}
\end{figure}

The velocity separating quasi-elastic and inelastic collisions, $c_{\min }$,
is shown as a function of $\chi $ in Fig. \ref{fig6}. Naturally, $c_{\min }$
increases with the decrease of $\chi $, as this implies the growth of the
amplitudes of the colliding solitons, and thus moving farther from the
integrable CNLSE\ limit, where collisions between solitons are completely
elastic.

We also simulated collisions between solitons with a phase shift of $\pi $
between them. In that case, as might be expected, the colliding solitons
always demonstrate a quasi-elastic rebound, irrespective of the values of $%
\chi $ and $c$ (not shown here).

Finally, three-soliton collisions (between two moving solitons and a central
quiescent one) were considered too. The results for them (not shown here)
are similar to those presented above for collisions between two solitons:
quasi-elastic passage of fast solitons, and the merger of slowly moving ones.

\section{Conclusion}

This work aimed to present a new physically relevant variant of the
NPSE (nonpolynomial Schr\"{o}dinger equation), which naturally
appears in the model of the RABR\ (resonantly absorbing Bragg
reflector), as the equation for gap solitons residing near the edge
of the bandgap. The equation features the nonlinearity in the form
of the radical term. Previously, different forms of the NPSE were
derived as models of the BEC\ dynamics in Refs. \cite{Luca} and
\cite{Vicente}. A full family of bright-soliton solutions to the
present NPSE was obtained in an implicit form, the ultimate solution
being the explicitly found ``quasi-peakon". Cnoidal waves in the
form of a chain of quasi-peakons were found too. The stability of
the bright solitons is correctly predicted by the VK criterion,
which separates the family into stable and unstable parts. The
ultimate form of the dark-soliton solution, shaped as a ``dark
compacton", was also obtained.

The instability has a peculiar form in the present model: it occurs when the
amplitude of the wave field attains the critical value ($|V|~=1$), which is
followed by the switch of the system into an unstable phase (which
corresponds to the inverted atomic population unsupported by the external
field), in terms of the underlying RABR model. The passage of the system
through the instability threshold was investigated analytically.

The dynamics of the stable bright solitons was further explored by means of
direct simulations. In particular, 2-soliton breathers were found, and the
border between the merger and mutual passage of colliding solitons was
identified. The merger may lead either to the formation of robust
irregularly oscillating pulses, or to the switch into the unstable phase.

\end{document}